# A New Synthesis Approach for Carbon Nitrides: Poly (Triazine Imide) and Its Photocatalytic Properties

Leonard Heymann[1], Björn Schiller[1], Heshmat Noei[2], Andreas Stierle[2,3], Christian Klinke[1,4,*]

[1] Institute of Physical Chemistry, University of Hamburg, Grindelallee 117, 20146 Hamburg

[2] DESY NanoLab, Deutsches Elektronen-Synchrotron DESY, 22607 Hamburg, Germany

[3] Physics Department, University of Hamburg, 20355 Hamburg, Germany

[4] Department of Chemistry, Swansea University, Singleton Park, Swansea SA2 8PP, United Kingdom

*Corresponding author: klinke@chemie.uni-hamburg.de

**Abstract**  Poly (triazine imide) (PTI) is a material belonging to the group of carbon nitrides and has shown to have competitive properties compared to melon or g-$C_3N_4$, especially in photocatalysis. As most of the carbon nitrides PTI is usually synthesized by thermal or hydrothermal approaches. We present and discuss an alternative synthesis for PTI which exhibits a pH dependent solubility in aqueous solutions. This synthesis is based on the formation of radicals during electrolysis of an aqueous melamine solution, coupling of resulting melamine radicals and the final formation of PTI. We applied different characterization techniques to identify PTI as the product of this reaction and report the first liquid state NMR experiments on a triazine-based carbon nitride. We show that PTI has a relatively high specific surface area and a pH dependent adsorption of charged molecules. This tunable adsorption has a significant influence on the photocatalytic properties of PTI which we investigated in dye degradation experiments.

# Introduction

Carbon nitrides have emerged as a new class of semi-conducting materials with a large variety of possible applications. Since photocatalytic water splitting has been reported for these compounds in 2009[1] most of them were prepared by thermal or hydrothermal synthesis leading in the formation of melon[2,3] (often referred as g-$C_3N_4$)[4-9], poly melamine[10-11], poly (heptazine imide) (PHI)[12-13] or poly (triazine imide) (PTI)[14-18]. Kessler *et al.* discuss in their review the differences of these structures and their promising properties, especially in photocatalysis.[19] In the literature it was shown that carbon nitrides are capable of reducing carbon dioxide[20-25], splitting water[1,20,26-29], degrading organic pollutants like dyes[6,12,22,30-32] under illumination in the visible range as well as having a high sensitivity in sensing applications[31,33-37] and catalyzing organic reactions.[38-42]

Different approaches have been used to improve the photocatalytic properties of these carbon nitrides. Ong *et al.* give a good overview over the so far published carbon nitrides and their modifications.[43] By elemental doping[44-47], copolymerization[48-49], templating[50-51] or forming composites especially with nanocrystals[52-57] it was possible to further extend the number of applications in photocatalysis, and even dark photocatalysis has been reported recently by Lau *et al.*[58] Being restricted to thermal and hydrothermal synthesis limits the variety in the above mentioned approaches. Therefore, it is important to discuss alternative strategies for the synthesis in order to further improve the properties of carbon nitrides.

In 2015 Lu *et al.* reported the synthesis of graphitic carbon nitride nanosheets using a new electrochemical approach and showed the peroxidase-like activity of those nanosheets.[59] In a new experimental set-up with differences in the subsequent washing procedure we were able to obtain a carbon nitride material either in solution (aqueous sodium hydroxide solution at ph 13 and DMSO) or as suspension (water at neutral pH) that can be freeze-dried yielding the powdery product. We applied different techniques in solid state like XPS, XRD and FTIR, but also liquid state techniques like two-dimensional 1H-15N-NMR spectroscopy to identify the product of the synthesis as poly (triazine imide). Reasonable yields could be achieved and the as-described synthesis is facile and safe compared to molten salt synthesis, which were conducted under high pressure.[14-18] Beside the thorough characterization, we were able to demonstrate and investigate PTIs adsorption behavior for charged molecules by the examples of naphthol yellow S (anionic) and methylene blue (cationic). Additionally, we demonstrated its capability in photocatalytic dye degradation of methylene blue and we further investigated the active species in these processes.

## Results and Discussion

The as-prepared product is proposed to be poly (triazine imide) (PTI). A schematic of the molecular structure is shown in Figure 1a. In the following passages different characterization methods are combined to confirm this hypothesis. In general, the yellowish brownish product can be obtained in solution or in suspension depending on the amount of sodium hydroxide in the synthesis. Figure 1b shows PTI in aqueous solutions at different pH values. The Tyndall-effect describes the scattering of an incident beam by small particles. Therefore, the beam is observable not only in the direction of propagation but also at different observation angles. The absence of this effect at pH 13 indicates the formation of a solution whereas at pH 1 and pH 7 a suspension is formed. Even after being synthesized and washed PTI can be either dissolved in aqueous solution due to the formation of the PTI anion as well as being again precipitated by protonation. We assume that the imide group can be deprotonated by the addition of a base, e.g. sodium hydroxide. The deprotonated and negatively charged PTI can be processed and analyzed in solution, which opens the possibility of a wide range of characterization techniques, e.g. liquid state NMR spectroscopy. The addition of an acid like hydrochloric acid results in the precipitation of PTI. This process does not change the observed properties of PTI. Besides, Miller *et al.* recently described the spontaneous dissolution of PTI in DMSO.[60] We were able to reproduce these results and observed a solubility of PTI at room temperature of 9.1 mg/mL and 10.2 mg/mL in DMSO and sodium hydroxide solution at pH 13, respectively. The solubility was determined by preparing a saturated solution of PTI in each of the solvents, filtering the supernatant with a syringe filter followed by the evaluation of the mass of PTI in a defined volume.

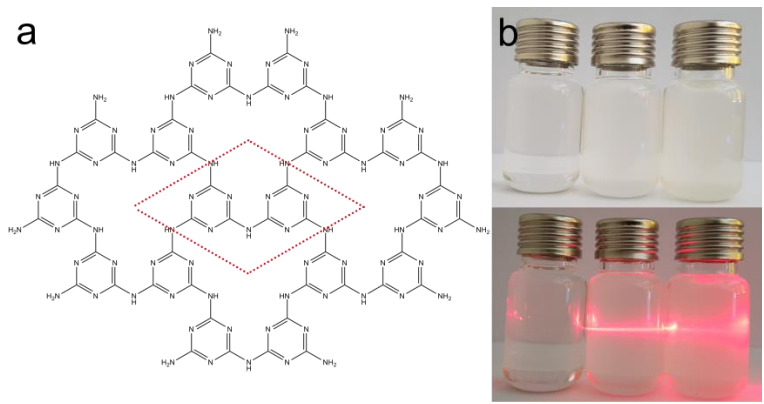

**Figure 1** The structure of a single plane of PTI and its unit cell (red) (a). Picture of PTI at different pH values (from left to right: pH 13, pH 1 and pH 7) (b). At pH 1 and pH 7 a nonstable suspension is obtained while at pH 13 a solution is formed (top). The lower image shows the Tyndall-effect at pH 1 and pH 7. At pH 13 the Tyndall-effect cannot be observed (bottom).

Lu *et al.* described the formation process of their graphitic carbon nitride as a coupling of oxidized melamine radicals.[59] These radicals were formed by the reaction of melamine with radical species emerging during the electrolysis of water, namely hydroxyl and superoxide radicals. These radicals are in-situ formed at voltages exceeding the standard potentials for electrolysis of water of 1.23 V plus the overpotential. A similar approach has been described for carbon nanostructures prepared using electrochemical exfoliation by Lu *et al.*[61] It can be observed that the colorless solution of melamine in water turns yellowish during the reaction. We noticed that this change of color takes place at the anode, which leads us to the assumption that radicals, which are evolving during the oxidation of water, may be the main reactive species in this reaction. To further prove that assumption we studied the formation of hydroxyl radicals. In order to do so terephthalic acid was dissolved in 1 M sodium hydroxide solution. The acid has been added next to the anode. Aliquots were taken after certain time intervals. The formation of 2-hydroxyterephthalic acid was observed by detecting the emission at 425 nm. The emission spectra are shown in Figure S2. The influence of the as-formed hydroxyl radicals was proven through conducting the regular experiment in the presence of terephthalic acid. It was observed that the color change appears later and less PTI is yielded.

**Characterization**

In the following section different methods are combined to identify PTI as the product of the above described reaction and its physical properties are discussed.

In order to determine the morphology of PTI transmission electron microscopy (TEM) and scanning electron microscopy (SEM) images were evaluated (Figure 2). Porous networks of PTI which are composed of loose ribbon-like fragments can be observed. These fragments are up to 500 nm long and up to 80 nm wide. In selected area electron diffraction (SAED) only broad rings with low intensity can be observed in very dense areas. Therefore, we assume that these PTI ribbons have small crystalline domains and are not single crystalline. The significant porosity of the networks is confirmed by the specific surface of 96 $m^2g^{-1}$ found by multipoint BET isotherm analysis (adsorption/desorption isotherms as well as a multipoint BET fit are shown in Figure S3) and is relatively high compared to other carbon nitride materials.[20,22,29]

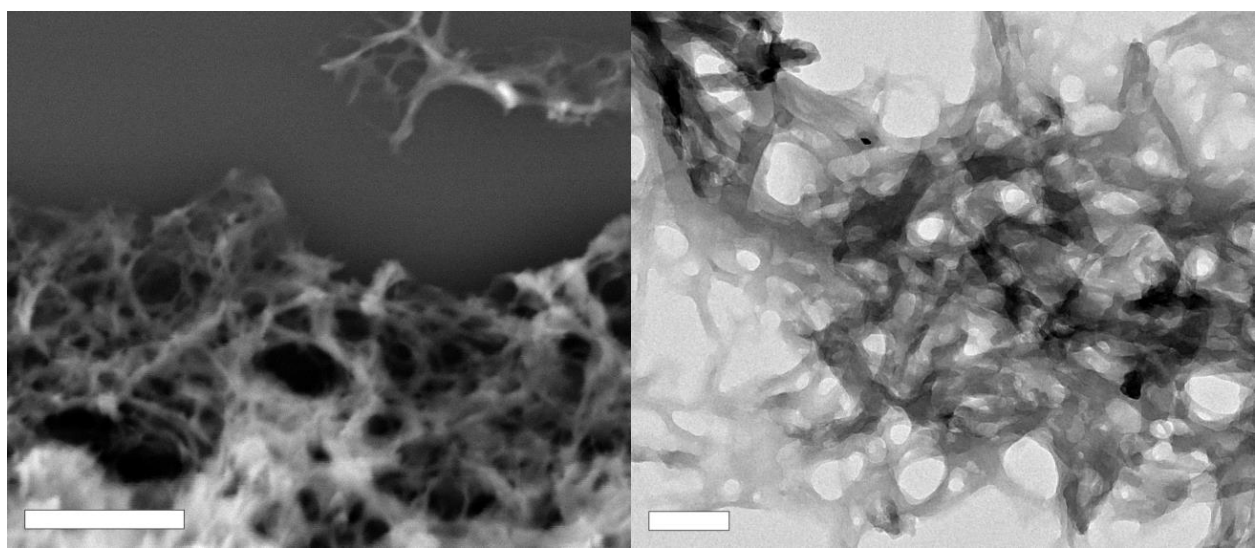

**Figure 2**     SEM and TEM images of the obtained PTI. The ribbon-like fragments arrange in loose networks with a high porosity. The scale bars in the SEM and TEM images are 500 nm and 100 nm, respectively.

When PTI is dissolved in DMSO a change in the morphology is observed. The SEM and TEM images are shown in Figure S4. The obtained structures exhibit a two-dimensional morphology.

Despite the above mentioned small domain sizes two characteristic reflexes can be observed in powder X-ray diffraction (XRD) of the as-obtained product. The reflex at 27.66 ° is related to the (002) stacking of conjugated aromatic systems. The calculated interlayer distance is found to be 3.22 Å. The values for tri-*s*-triazine based carbon nitride are 27.4 ° and 3.26 Å, respectively.[1] The less intense reflex at 10.6 ° is labeled as (100) and is defined by the distance of 8.33 Å. This distance is found in the repeating unit in plane, more precisely in the periodicity of the pores formed by the *s*-triazine subunits. The shift in the reflex of the *s*-triazine based carbon nitride compared to the tri-*s*-triazine based carbon nitride (13.0 ° and 6.81 Å)[1] is clearly observable and is explained through the smaller unit cell of the latter. Figure 3a shows the XRD pattern of the proposed *s*-triazine based structure shown in Figure 1. The distance of 8.33 Å shows a slight shift compared to values found for PTIs with intercalated metal ions synthesized using hydrothermal approaches.[15,62,63] The broadening of the reflexes indicates small crystal sizes. Using the Scherrer equation a crystal size of around 9 nm is found for the analysis of the (100) reflex. The crystal size found for the reflex (002) at 27.66 Å is 14 nm. We explain the absence of higher order reflexes by the size of the crystals, which is low compared to PTI prepared by molten salt synthesis. Wirnhier *et al*. report sizes of up to 60 nm.[14]

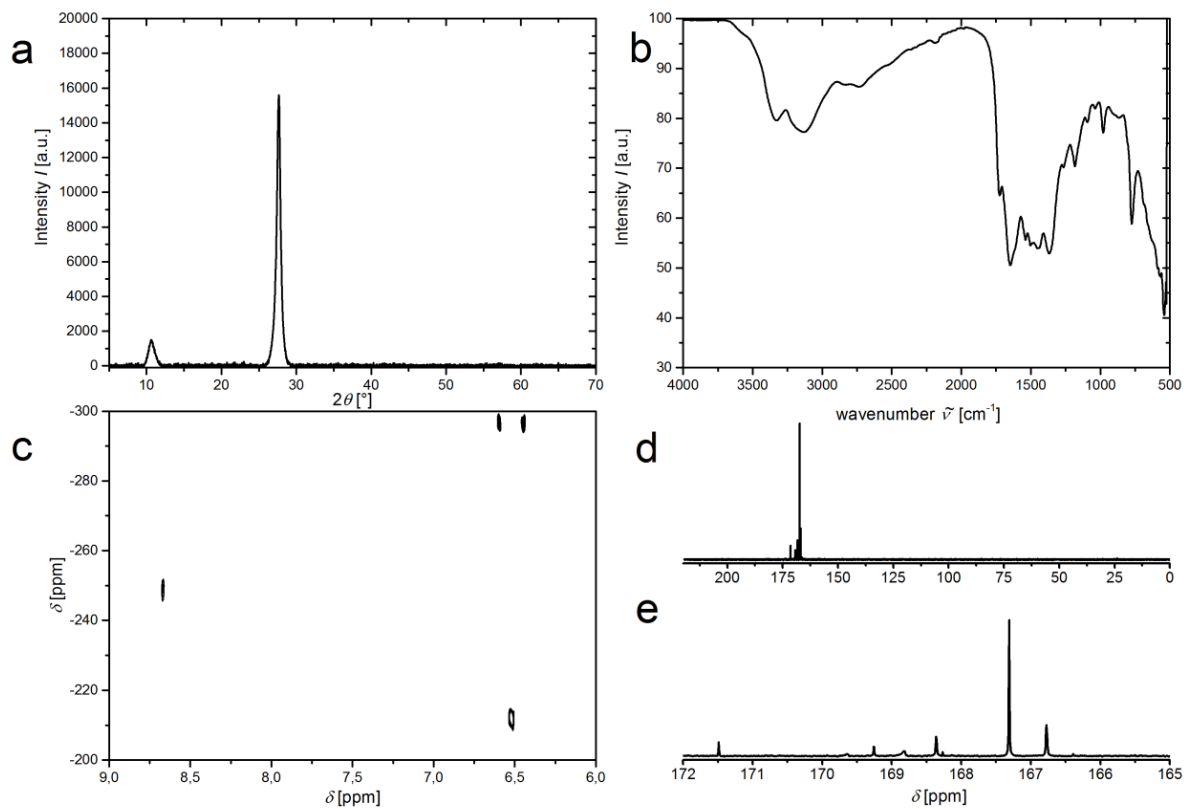

**Figure 3** The XRD pattern of PTI shows two characteristic reflexes at 10.6 ° and 27.4 °, which are related to an in-plane periodicity (100) and interlayer stacking (002), respectively (a). FTIR spectrum of PTI recorded between 500 cm$^{-1}$ and 4000 cm$^{-1}$. The spectrum shows characteristic absorptions at 775 cm$^{-1}$, between 1100 cm$^{-1}$ and 1700 cm$^{-1}$, especially at 1260 cm$^{-1}$ and around 3130 cm$^{-1}$ and 3330 cm$^{-1}$(b). 1H, 15N HMBC NMR spectrum of PTI in DMSO. Signals are observed at -212.2 ppm, -249.5 ppm and -296.4 ppm in respect to nitromethane. These signals can be assigned to the central ring nitrogen and the bridging imide and amine nitrogen, respectively (c). 13C-NMR spectrum of PTI in DMSO. Signals are observed between 166.7 ppm and 171.5 ppm. These peaks can be assigned to the carbon atoms in triazine and heptazine rings (d and e).

In order to confirm our hypothesis the elemental composition was evaluated through combustion (carbon, nitrogen and hydrogen) and pyrolysis analysis (oxygen). The determined values are presented in Table 1.

**Table 1**    Amounts of carbon, nitrogen, hydrogen and oxygen found in the sample by combustion and pyrolysis analysis.

|  | C  mass% (atomic%) | N  mass% (atomic%) | H  mass% (atomic%) | O  mass% (atomic%) |
|---|---|---|---|---|
| Combustion analysis | 26.11 (20.84) | 47.79 (32.71) | 3.67 (34.91) | / |
| Pyrolysis analysis | / | / | / | 19.27 (11.54) |

For an integer amount of carbon a sum formula of $C_3N_{4.7}O_{1.7}H_5$ is calculated. These values are in good agreement with the values of $C_3N_{4.5}H_{1.5}$ given for the unit cell of PTI defined in Figure 1a. The nonconformity of the values found for hydrogen and oxygen can be explained by adsorbed water and for hydrogen by primary amine groups at the border of the structures or vacancies. A significant amount of oxygen can also be found in functional groups proven using FTIR and XPS measurements.

Figure 3b shows the FTIR spectrum of the product. There are multiple bands, which can be assigned to different absorptions which are characteristic for carbon nitrides in general. The sharp peak at 775 cm$^{-1}$ can be ascribed to the out-of-plane bending mode of either triazine or heptazine rings. The fingerprint region between 1100 cm$^{-1}$ and 1700 cm$^{-1}$ shows various absorption maxima for imide and nitride stretching vibrations. Around 1260 cm$^{-1}$ the characteristic band for the C-NH-C unit is found. The bands around 2170 cm$^{-1}$, 1720 cm$^{-1}$ and 1180 cm$^{-1}$ indicate the presence of nitrile or oxygen containing functional groups like C=O or C-OH. The bands at 3330 cm$^{-1}$ and 3130 cm$^{-1}$ are related to NH-stretching vibrations.[14,15] To further confirm PTI as the product of the as-described synthesis nuclear magnetic resonance (NMR) and X-ray photoelectron spectroscopy (XPS) experiments were conducted.

NMR spectroscopy is a powerful tool to identify the chemical structure of PTI and to exclude a heptazine-based structure. Up to now most carbon nitrides were analyzed by solid state MAS-NMR spectroscopy[2,14,48,59,63] due to the vanishing solubility of carbon nitrides in common solvents. Unfortunately, compared to liquid state NMR spectroscopy this method is time-consuming and lacks resolution. For polymeric carbon nitride (melon), a heptazine-based carbon nitride, NMR spectra have been recorded in concentrated sulfuric acid.[64]

The 13C-NMR spectrum of PTI shown in Figure 3d-e is defined by several signals distributed in the region between 166.7 ppm and 171.5 ppm. Signals in that region are typically observed for both triazine and heptazine based carbon nitride materials.[2,14,48,59,63] Therefore, the 15N-NMR spectrum is needed to assign the observed 13C-NMR signals to the triazine based PTI. The carbon atoms in the triazine ring are located next to imide groups as well as primary and secondary amine groups. Also nitrile or oxygen containing functional groups are observed in FTIR and XPS spectra and contribute to the shown NMR spectrum by slight shifts in this region. Figure S5a shows the solid state CPMAS 13C NMR spectrum of PTI. It consists of two broad signals at 165.5 ppm and 157.2 ppm. The observed shifts result by the stacking of the PTI planes in solid state. Also solvent interaction has to be reconsidered. The 15N-NMR spectrum of PTI was collected in a heteronuclear multiband correlation experiment (Figure 3c). It consists of three signals at -212.4 ppm, -249.5 ppm and -296.8 ppm. The first signal is assigned to the tertiary nitrogen atom in the outer ring of triazine or heptazine units in accordance to literature.[14,15] The absence of a signal reported for heptazines in the region of -225 ppm to -235 ppm suggests that no central heptazine nitrogen atom is present.[2] The signal at -249.5 ppm can be assigned to bridging NH groups of PTI. The latter signal at -296.4 ppm is ascribed to the nitrogen atom in the amine groups. These assignments are confirmed by the correlated 1H-NMR spectrum, which shows three signals for the amine group centered at 6.52 ppm and one signal at 8.68 ppm for bridging imide groups. Figure S5b shows the solid state CPMAS 15N NMR spectrum of PTI. Three main signals at -218.0 ppm, -241.5 ppm and -292.6 ppm are present and confirm the liquid state NMR.

X-ray photoelectron spectroscopy (XPS) measurements are performed to analyze the PTI composition. Figure S6a shows the survey spectra of the PTI and demonstrate the corresponding C 1s and N1s peaks. The deconvoluted XP spectra of C1s and N1s regions are shown in Figure 4. The deconvoluted C1s peak (Figure 4a) demonstrates four components at binding energies (BE) of 284.5, 285.3, 288.15, 289.5 eV. These components correspond to the graphitic carbon, aliphatic carbon, C-N, and carboxylic groups, respectively. The dominant C1s peak at 288.1 eV originates from sp2, where carbon atom bonds to N inside the triazine ring.[65,66] The contribution of the carbon groups at binding energy of 289.5 eV are attributed to the hydroxyl and carboxylate groups and reveals oxidation of the surface during the electrochemical induced synthesis. The corresponding O peaks for the –OH and –COOH are observed at 531.1 and 532.1 eV, respectively (see O1s peak in Figure S6b). Deconvoluted N1s spectra presented in Figure 4b shows three peaks centered at 398.4, 399.6 and 400.7 eV assigned to the sp2 nitrogen in the triazine ring, bridging N atoms and the primary amine groups, respectively.[12,65-67]

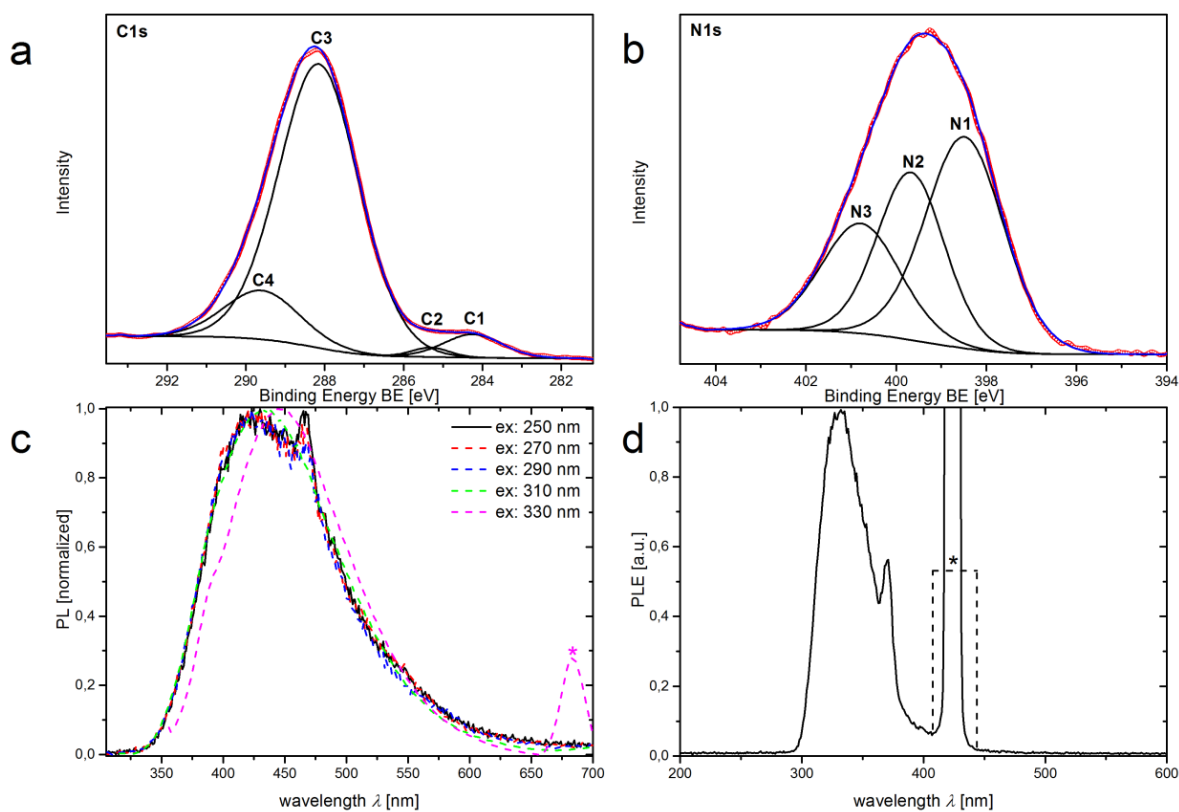

**Figure 4** C1s (a) and N1s (b) XP spectra of PTI. Red circles denote the measured data and the fits are represented by blue curves. With an increasing excitation wavelength the photoluminescence spectra of PTI in NaOH solution (pH 12) shows a redshift in the photoluminescence (c). The PLE spectrum shows two maxima. The first and more intense is found at 330 nm and is contributed by PTI, the second at 370 nm is assigned to Raman scattering of water (d). The asterisks in (c) and (d) mark measurement artefacts at double excitation wavelength and the excitation wavelength, respectively.

Thermogravimetric analysis shows the high temperature stability of PTI (Figure S7). With an increasing temperature water desorbs and at 100 °C a plateau is reached independently of the surrounding atmosphere. At 300 °C in an ambient atmosphere a constant mass loss is observed, which we assume to be determined by the sublimation of PTI as well as the complete decomposition at higher temperatures. A complete mass loss is observed at 575 °C. In a nitrogen atmosphere at 300 °C significant mass loss is observed as well, but a second plateau is reached at 450 °C. This plateau may be reached due to further condensation of PTI. It is followed by further mass loss until a minimum of 2% residue mass is observed at 675 °C. In contrast to the measurements in ambient atmosphere a black residue remains.

The photoluminescence (PL) spectrum of PTI in NaOH at pH 12 indicates a distinct emission maximum at 425 nm at an excitation wavelength of 250 nm (Figure 4a). With an increasing excitation wavelength a redshift is observed up to 447 nm. The redshift might be explained by electronic transitions which become unlikely at lower energies and are in general observed due to the changes in the electronic structure of PTI due to the deprotonation. This property is not observed for PTI in DMSO (shown in Figure S8a). The emission maximum only slightly varies around 443 nm. In both solvents the emission covers a large range of the visible spectrum, which makes PTI a promising material for UV and white light LEDs. This property as well as the spontaneous dissolution of PTI in polar aprotic solvents (DMF, DMSO, NMP, Dimethylacetamide) has been recently described by Miller et al.[60] In contrast to their reports, we do not observe any broadening of the emission with an increasing excitation wavelength of PTI neither in NaOH nor in DMSO. The fwhm found for the latter is around 120 nm independently of the excitation wavelength, while the former shows a narrower fwhm of around 100 nm. The photoluminescence excitation (PLE) spectrum of PTI in NaOH solution shows a high intensity at 332 nm, as it has also been reported by Miller et al. Additionally, we observe a smaller maximum at 370 nm whereas the reported maximum of few layer PTI nanosheets at 270 nm could not be observed in NaOH solution (Figure 8b). Raman scattering of water results in the maximum at 370 nm. In DMSO the PLE maximum is found at 357 nm and only one peak is observed (Figure S8b). We explain the differences in the spectra by emphasizing the different mechanism of the dissolution processes. In NaOH PTI is deprotonated and therefore, the electronic properties are influenced whereas in DMSO PTI is not chemically altered. The energy gap between HOMO and LUMO of 3.28 eV is determined through a Tauc plot (Figure S9b), which was extracted from a diffuse reflectance spectrum (Figure S9a).

**Photocatalytic experiments**

Prior to photocatalytic experiments the adsorption of ionic dyes on PTI was investigated. The cationic dye methylene blue (MB) and the anionic dye naphthol yellow S (NYS) were used. Due to charges generated by adjusting the pH value the adsorption behavior of the ionic dyes is influenced. It was shown that MB adsorbs well at neutral (pH 7, capacity for MB: 0.04 mmol/g) and increased pH values (pH 11, capacity for MB: 0.06 mmol/g), but has a small adsorption capacity at low pH values (pH 3, capacity for MB: 0.02 mmol/g). For NYS in contrast the adsorption at neutral (pH 7, capacity for NYS: 0.07 mmol/g) and increased pH values (pH 11, capacity for NYS: 0.03 mmol/g) is small compared to decreased pH values (pH 3, capacity for NYS: 0.22 mmol/g). Hence, we show that by adjusting the pH value in the experiment specific adsorption of charged molecules can be achieved. This property is important for the selectivity of the catalyst to bind specific molecules and influences the catalytic activity. Figure 5a shows the pH value dependent adsorption capacity of PTI for MB and NYS. The adsorption capacity was determined at the equilibrium of the adsorption and desorption processes, which was reached after 60 min (Figure S10).

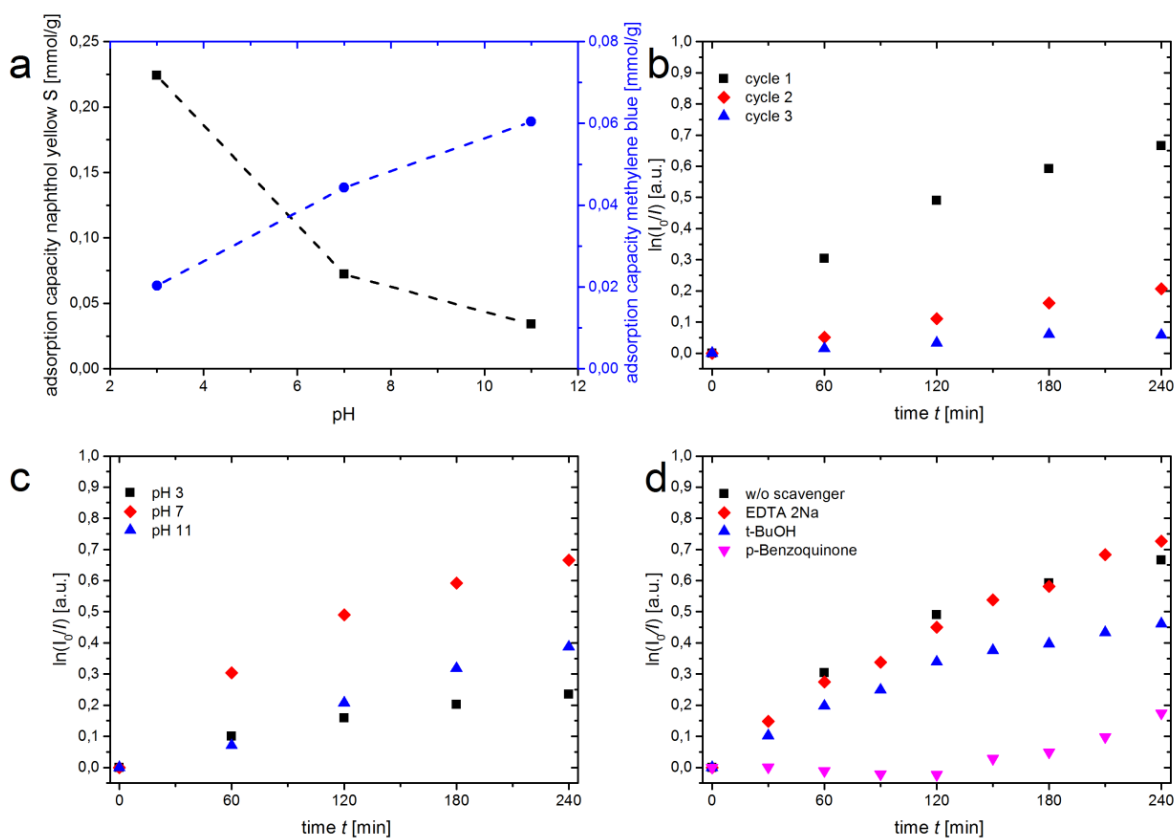

**Figure 5** Adsorption capacity of PTI for the ionic dyes NYS (black squares) and MB (blue circles) at different pH values. PTI shows a higher capacity for the anionic NYS at low pH values due to charging of PTI by protonation. In contrast, the adsorption of cationic MB is increased at high pH values because of the charges formed due to deprotonation of PTI (a). Logarithmic plots of the intensity of the MB solution against the duration of the degradation in recycling experiments (b) as well as at different pH values (c). The intensity was determined at the absorption maximum at 664 nm. The reaction constants were determined by fitting the linear region at 120 to 240 min. Logarithmic plot of intensity of the MB solution against the duration of the degradation at pH 7. Different scavengers were used to determine the active

species in the degradation process. It is shown that *p*-Benzoquinone lowers the degradation rate for a certain time (d).

These experiments were followed by dye degradation experiments under illumination with a xenon lamp in combination with a water filter to exclude thermal effects. For NYS no significant degradation was observed, therefore only the degradation of MB is described in the following section. The spectrum of the lamp used for the photocatalytic experiments is shown in Figure S11. The UV-Vis spectra used to determine the intensity at the maximum at 664 nm as well as the references are shown in the Supporting Information (Figure S12).

**Degradation of methylene blue**

The performed experiments proof the photocatalytic activity of PTI in the degradation process of MB. It was shown that degradation of MB could be observed for at least three cycles at neutral pH value (Figure 5b). However, it has to be conceded that the degradation rate decreases significantly by a factor of 3 (cycle 1: $k_1$=1.5·$10^{-3}$min$^{-1}$, cycle 2: $k_2$=0.8·$10^{-3}$min$^{-1}$ and cycle 3: $k_3$=0.4·$10^{-3}$min$^{-1}$, $k$ was determined using linear regression of the data points for 120, 180 and 240 min). Besides, it has also been shown that the pH value has a strong influence on the reaction rate constant. The logarithmic plots are shown in Figure 5c. At low pH values the adsorption of MB is lowered (0.02 mmol/g at pH 3), whereas at high pH values (0.06 mmol/g at pH 11) a three times higher amount of dye can be adsorbed. The degradation rates determined at pH 3, pH 7 and pH 11 are $k_{pH\ 3}$=0.6·$10^{-3}$min$^{-1}$, $k_{pH\ 7}$=1.5·$10^{-3}$min$^{-1}$ and $k_{pH\ 11}$=1.5·$10^{-3}$min$^{-1}$. It can be observed that by adjusting the pH the degradation rate can be influenced. In comparison to the degradation rates at pH 7 and pH 11 the rate at pH 3 is lowered by the factor of 2.5 which is in good agreement with the observed lower adsorption capacity.

**Investigation of different active species**

Different scavenger molecules have been added to the photocatalytic experiments to investigate the active species in the degradation process. The influence of holes, hydroxyl radicals as well as superoxide radicals was determined by adding EDTA-2Na, *t*-Butanol and *p*-Benzoquinone, respectively. As shown in Figure 5d at pH 7 the addition of EDTA-2Na does not have any influence on the degradation rate ($k_{EDTA-2Na}$=2.3·$10^{-3}$min$^{-1}$) of MB. Thus, it is assumed that hydroxyl radicals are not the active species in the degradation process. Adding *t*-BuOH to the experiment results in a slightly lowered degradation rate ($k_{t-BuOH}$=1.0·$10^{-3}$min$^{-1}$). In contrast, *p*-Benzoquinone has a significant influence. It is observed that for nearly 2 hours of illumination no degradation is observed and no reaction constant could be determined. Instead, a shift in the absorption of *p*-Benzoquinone is observed. After 2 hours no further shifting is occurring and the degradation of MB takes place. The reaction constant in this regime is in the same range as the degradation without any scavenger. Therefore, it is concluded that superoxide radicals are the main active species and that these radicals only react with MB after the reaction with *p*-Benzoquinone is completed.

## Conclusion

In summary, this work describes a new approach for the synthesis of a triazine based carbon nitride material which we identified as poly (triazine imide). This approach has its origin in a synthesis described for triazine based graphitic carbon nitride by Lu *et al.*[59] Our modification of the synthesis approach yields PTI, which we identified with the help of several characterization techniques, e.g. XPS, powder XRD, FTIR and two dimensional liquid state NMR spectroscopy. Latter is a rarity for carbon nitride materials because the solubility in common solvents is low. Due to this synthesis approach oxygen containing functional groups are incorporated in the structure of PTI resulting in the possibility of protonation and deprotonation. This charging leads to an increased solubility. The as-obtained PTI has a comparably high specific surface area and also shows high temperature stability. We showed the photocatalytic properties of PTI in the degradation process of ionic dyes and discussed the influence of pH value in this process, especially in the adsorption process. We were able to influence the adsorption capacity of PTI for different molecules by adjusting the pH value. The addition of different scavengers showed that superoxide radicals are the main active species in the presented degradation experiments.

## Experimental Section

### Chemicals

Melamine (99 %) was purchased from Sigma-Aldrich, sodium hydroxide (97 %) was purchased from Grüssing GmbH, methylene blue was purchased from Merck Chemicals GmbH, naphthol yellow S was purchased from Alfa Aesar, *tert*-Butanol (99 %) was purchased from Grüssing GmbH, EDTA-2Na (>98.5 %) was purchased from Sigma-Aldrich, *para*-Benzoquinone was purchased from Sigma-Aldrich and recrystallized in 2-propanol, terephthalic acid (98 %) was purchased from Sigma-Aldrich.

**Synthesis of poly (triazine imide)**

945 mg (7.5 mmol) of melamine were dissolved in 100 mL of demineralized water at 65 °C under vigorous stirring. In a three electrode set-up with a reference electrode (Ag/AgCl/KCl), counter electrode (Platinum) and working electrode (Platinum) a voltage of 5 V was applied for 6 hours at 65 °C without stirring. The distance between counter and working electrode was 0.5 cm. The current increased linearly during the reaction from 1 mA to 200 mA and the formation of gases at the electrode surfaces increased as well. After 6 hours the product was obtained in a brownish suspension. A scheme of the set-up is shown in Figure S1.

The suspension was centrifuged for 5 min at 7500 rpm (13206 rcf). The supernatant was discarded. The residue was dispersed in 30 mL of hot demineralized water. This process was repeated three times. The washed product was freeze dried and obtained as a pale yellowish powder. This synthesis approach and the following washing procedure yield in average 55 mg of PTI.

**Photocatalytic experiments**

50 mg of the catalyst material were dispersed in 100 mL of either a 10 mg/L methylene blue or a 60 mg/L naphthol yellow S solution and stirred for one hour in the dark to set an adsorption/desorption equilibrium. As a light source a 300 W xenon lamp (UXL-302-O) combined with a water filter to avoid heating effects by infrared radiation was used. Aliquots of 4 mL were taken every 60 min (or 30 min for scavenger experiments). The aliquots were centrifuged once at 11000 rpm (28408 rcf) in order to remove the catalyst material. The supernatant was filtered using a hydrophilic syringe filter (nylon, 0.2 µm pore size).

The pH value was adjusted in the experiments by the addition of hydrochloric acid or sodium hydroxide before the addition of the catalyst. All of the measurements were controlled by performing the same experiments without illumination as well as without the catalyst.

In order to identify the reactive species in the degradation process experiments were conducted as described above, but *tert*-Butanol (10 vol.%), EDTA-2Na (10 mM) or *para*-Benzoquinone (10 mM) were added to the experiment.

## Associated Content

**Supporting Information**

The Supporting Information is available free of charge on the ACS Publications website at DOI: .

Experimental methods, Scheme of the experimental set-up, formation of hydroxyl radicals during the reaction, BET measurements, SEM and TEM images, solid state 13C and 15N NMR spectra, full XP spectrum and deconvoluted O1S peak, TGA of PTI in ambient and nitrogen atmosphere, UV-Vis spectra of PTI in solid state, resulting tauc plot and UV-Vis spectra in DMSO, xenon lamp spectrum, and UV-Vis spectra of aliquots taken during dye degradation.

## Author Information

**Corresponding author**

*E-mail: klinke@chemie.uni-hamburg.de

**ORCID**

Leonard Heymann: 0000-0001-6177-6943

Christian Klinke: 0000-0001-8558-7389

**Notes**

The authors declare no competing financial interest.

## Acknowledgement

L.H. and C.K. gratefully acknowledge financial support of the European Research Council via the ERC Starting Grant "2D-SYNETRA" (Seventh Framework Program FP7, Project: 304980). C.K. thanks the German Research Foundation DFG for financial support in the frame of the Cluster of Excellence "Center of ultrafast imaging CUI" and the Heisenberg scholarship KL 1453/9-2.